\documentclass[12pt,preprint]{aastex}

\usepackage[]{times, graphicx}
\newcommand{\brr}{$\langle B_{||} \rangle_{r=20}$}
\newcommand{\soho}{{\em SOHO{}}}
\newcommand{\pref}{\protect\ref}
\begin{document}

\shorttitle{Active Region Dynamics}
\shortauthors{S.~W. McIntosh}
\title{Does High[Plasma]\--$\beta$ Dynamics ``Load'' Active Regions?}
\author{Scott W. McIntosh\altaffilmark{1,2}}
\altaffiltext{1}{Department of Space Studies, Southwest Research Institute, 1050 Walnut St, Suite 400, Boulder, CO 80302}
\altaffiltext{2}{Visitor at the High Altitude Observatory,
National Center for Atmospheric Research, P.O. Box 3000, Boulder, CO 80307}
\email{mcintosh@boulder.swri.edu}

\begin{abstract}
Using long-duration observations in the \ion{He}{2}~304\AA{} passband of the Solar and Heliospheric Observatory (SOHO) Extreme-ultraviolet Imaging Telescope (EIT) we investigate the spatial and temporal appearance of impulsive intensity fluctuations in the pixel light curves. These passband intensity fluctuations come from plasma emitting in the chromosphere, transition region and lowest portions of the corona. We see that they are spatially tied to the supergranular scale and that their rate of occurrence is tied to the unsigned imbalance of the magnetic field in which they are observed. The signature of the fluctuations (in space and time) is consistent with their creation by magnetoconvection forced reconnection that is driven by the flow field in the high-$\beta$ plasma. The signature of the intensity fluctuations around an active region suggest that the bulk of the mass and energy supplied into the active region complex observed in the hotter coronal plasma is supplied by this process, dynamically forcing the looped structure from beneath. %This study augments contemporary analyses of quiet sun and coronal hole regions that have already established a link between spectroscopic diagnostics and energy input to the atmosphere by the relentless magnetoconvective forcing of the lower atmosphere. 
\end{abstract}


\section{Introduction}
Recent results \cite[][]{McIntosh2006a, McIntosh2006b} have provided detailed observational support for the (widely held) hypothesis that the relentless action of magnetoconvection-driven reconnection \cite[e.g.,][]{Priest2002} supplies the bulk of the energy to the solar atmosphere through ejecta that are intrinsically tied to supergranular spatial scales. In discussing their results, derived from Solar and Heliospheric Observatory \cite[\soho,][]{Fleck1995} observations, \cite{McIntosh2006b} have proposed that the small-scale eruptive phenomena observed are linked to the ubiquitous appearance of ``spicules'' at the solar limb in eclipse or coronagraph observations \cite[e.g.,][]{Secchi1877, Roberts1945}. The work presented in this paper was influenced by a movie presented on the \soho{} internet site\footnote{The movie can be viewed at {\url http://sohowww.nascom.nasa.gov/pickoftheweek/old/24mar2006/}.} from the Extreme-ultraviolet Imaging Telescope \cite[EIT;][]{Boudine1995} and shows the Sun ``percolating'' with passband intensity brightenings in the chromosphere and transition region that occur on spatial scales, commensurate with those discussed by \cite[][]{McIntosh2006b}. This Letter (and a subsequent Paper that is in preparation \-- McIntosh, Judge \& Gurman 2007) sets out to test the hypothesis presented by \cite[][]{McIntosh2006b} with the EIT data used to build this movie and several others like it while offering an interesting, alternative, but logical look at a possible source of energy and mass input of active regions by the relentless forcing of emerging (and existing) magnetic flux by the convective flow of the high-$\beta$ plasma.

In the following section, we will discuss the time-series observations from the 304\AA{} (\ion{He}{2}) passband of EIT that form the basis of our study. In Sect.~3 we will investigate the spatial and statistical dependence of the \ion{He}{2} 304\AA{} intensity light-curves to identify (and count) individual ``events'' and study the variance of the intensity in each spatial pixel over an appropriate period of time. In Sect. 4 we will demonstrate that the appearance of spatial patterns in the analysis of the 304\AA{} observations are specifically tied to the plasmaÕs magnetic environment, the chromospheric supergranular network and thus to the relentless action of magnetoconvection: emerging, shuffling and destroying small-scale magnetic flux elements that form the ``Magnetic Carpet'' \cite[e.g,][]{Schrijver1997}. We deduce, from the near ubiquity of the process, that its role is not restricted to loading mass and energy into the quiet sun and coronal holes, but is a significant source of the small-scale dynamics that are responsible for mass and energy \cite[with help from the mechanism discussed by][]{Jefferies2006} into active regions; in large part the thesis of \cite{Aschwanden2007}.

\section{Observations}\label{secobs}
The analysis presented in this Letter is based on the analysis of a long-duration sequence of EIT 304\AA{} passband imaging, starting 2006/03/18 20:36UT and ending 2006/03/21 23:48UT. The subset of the EIT sequence analyzed here (2006/03/19 00:11 \-- 2006/03/19 23:59) has a fixed cadence of twelve minutes and has no missing data blocks and covers the approximate evolutionary lifetime of a supergranular cell ($\sim$24 hours). We assume that the bulk of the EIT 304\AA{} passband consists primarily of emission from \ion{He}{2} formed in the upper chromosphere or low solar transition region at a temperature of $\sim$50,000~K \cite[][]{Mazzotta1998}. However, aside from the strong emission from the \ion{He}{2} 303.78\AA{} line there is a contribution to the bandpass emission in an active region from the \ion{Si}{11} 303.31\AA{} line and the blended emission of \ion{Mn}{14} and \ion{Fe}{15} at 304.86\AA{} (all three contaminant lines are formed in the hotter corona at $\sim$1.5MK) in the ratio 10::1::0.06 \cite[][]{Brosius1998}. The contribution of \ion{He}{2} emission relative to the emission lines of \ion{Fe}{10}/\ion{Fe}{11} (i.e., the contribution from the EIT 171\AA{} passband in second spectral order at about $\sim$1MK) is 25 to 1 in an active region and 10 to 1 in the quiet sun \cite[][]{Brosius1998}.

The 304\AA{} sequence is augmented by coronal context images from the EIT 195\AA{} passband and full disk line-of-sight magnetograms ($B_{||}$) from the Michelson Doppler Imager \cite[][]{Scherrer1995} closest to the start of the sequence, 01:12UT and 01:36UT respectively. Figure~\pref{march2006} shows the context images and subfield pointing for the 2006 March EIT sequence. In panels A through D of each figure we show the EIT 195\AA{} coronal image, the first EIT 304\AA{} image in the sequence, the MDI $B_{||}$ image and a smoothed $B_{||}$ image \cite[\brr;][]{McIntosh2006a} providing information about the net-magnetic flux balance on the supergranular scale (i.e., the magnetogram is convolved with circular kernel of diameter 20Mm). We also show the 1200\arcsec x 1200\arcsec{} field of view studied (black square in panels A and B, white in panels C and D). The thin black contour in panel D shows where \brr=0~G, the magnetic ``neutral'' line occurs.

The EIT data sample presented (reduced using the standard package discussed in the EIT User's Guide \-- {\url http://umbra.nascom.nasa.gov/eit/}) shows a selected subfield of the EIT field-of-view that has been tracked, extracted and co-aligned (to within half of a pixel) using the Interactive Data Language solar data mapping software in the SolarSoft data analysis tree \cite{Freeland1998}. %We estimate \cite[using a cross-correlation algorithm used to obtain sub-pixel timeseries co-alignment with TRACE data, e.g.,][]{McIntosh2004} that the data are tracked and co-aligned to within half of a pixel over the period needed, using the first image as a reference point.

\section{Data Analysis}\label{data}
The analysis of the EIT 304\AA{} data presented is very straightforward, is easily repeatable and has two parallel threads to explore the temporal and spatial behavior of the observations. In the first thread, we analyze the intensity light curves at each spatial pixel in the de-rotated, co-aligned and trend (a ten timestep box-car smoothed profile of the intensity) removed ($\hat{I}(x,y,t) = {I}(x,y,t) - <{I}(x,y,t)>_{10}$) datacube to count the number of significant brightenings over the observing period. Each brightening counts as one instance of an ``event'' and is defined as a 10\% change in the trend-removed ratio $\Delta \hat{I} /  \hat{I}$ (= [$\hat{I}(x,y,t-1) - \hat{I}(x,y,t)] / \hat{I}(x,y,t)$). The second thread investigates the spatial dependence of the brightenings through the pixel-by-pixel intensity distribution, $f(I(x \pm \delta x, y \pm \delta y,t))$, where $\delta x$ and $\delta y$ are each two pixels. Using a slightly extended spatial range ensures a higher signal-to-noise level in the time series (five times the number of points) and ensures that the moments of the intensity distribution, or a curve fit, can be accurately estimated with a high degree of reliability. The accumulation of data explicitly assumes that the plasma in the adjoining pixels is (in some way) physically coupled to that in the pixel being studied.

In Fig.~\pref{diag} we take an example pixel from the 2006 March sequence to demonstrate the quantities used in the analysis. Panel A shows the intensity timeseries data for the pixel over 20 hours (triangles) and the ten pixel boxcar trend (gray solid line). In panel~B we show the variation of $\Delta \hat{I} /  \hat{I}$ over the time period as well as the 10\% increase (dot-dashed) line we use to define occurence an ``event'', the total number of events in that pixel is then computed as half the number of times the $\Delta \hat{I} /  \hat{I}$ profile crosses the 10\% increase line. In panel~C we show the distribution of intensities for the pixel range, a Gaussian fit (thick gray line) to the distribution and the four moments (mean, standard deviation, skewness and kurtosis) of the distribution. We note that the use of skewness and kurtosis in this analysis is very much dependent on having a large enough number of samples to get an accurate intensity distribution \cite[see, e.g.,][]{Abramowitz1972,Press1992}. While they are not commonly used they serve a purpose in the present analysis. We use a simple definition of skewness: it measures whether outliers in the intensity distribution are below (negative) or above (postive) the distribution mean, skewing it to the left or the right respectively. Similarly, the kurtosis indicates whether the distribution is more or less sharply peaked than a Gaussian (a kurtosis of zero is perfectly Gaussian) while a flat value has a negative kurtosis and very sharply peaked distributions result in a positive kurtosis.

\section{Results}\label{results}
\begin{figure}
\epsscale{0.75}
\plotone{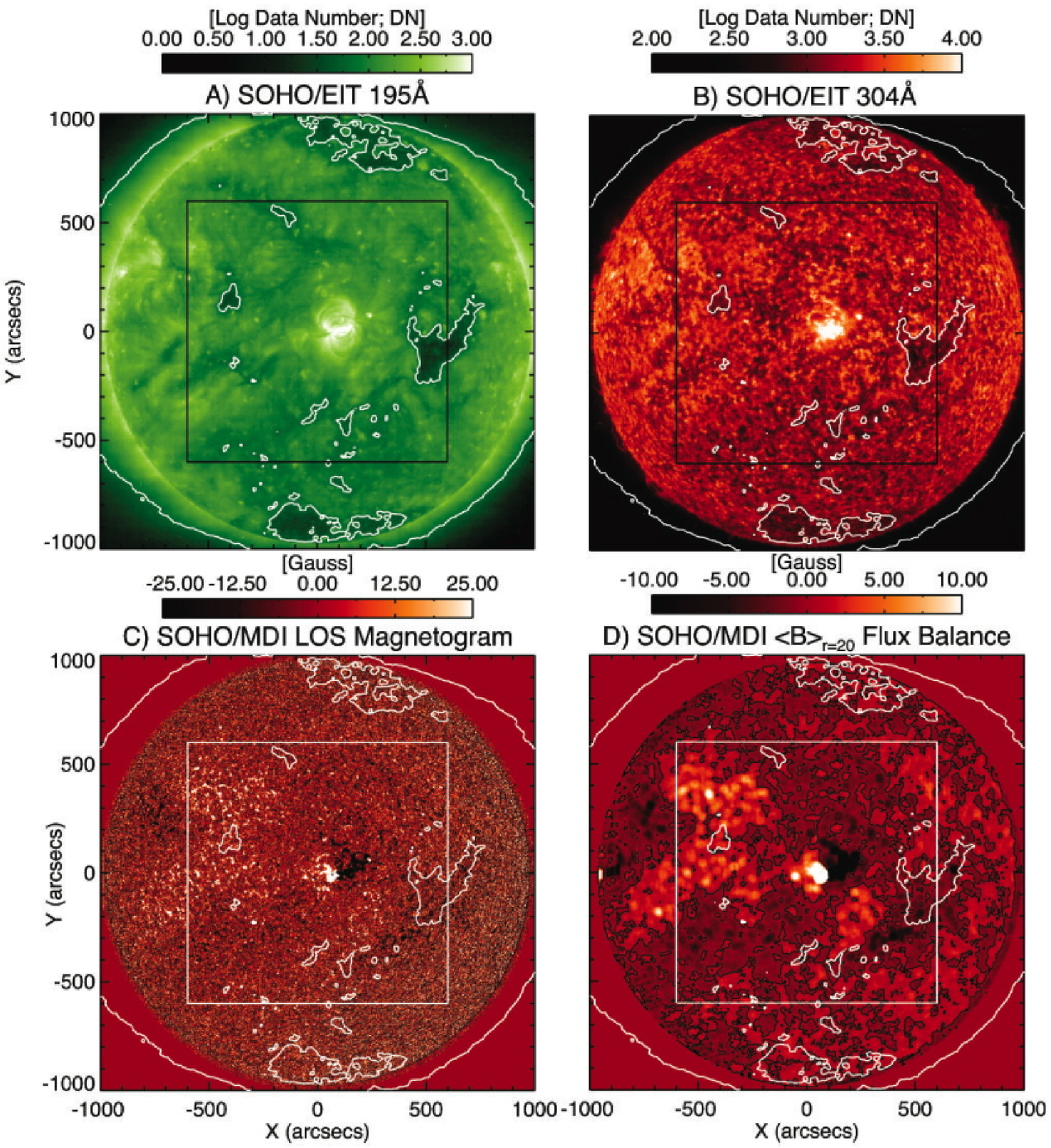}
\caption{Context images for the 2006 March 18 EIT 304\AA{} sequence. \label{march2006}}
\end{figure}

\begin{figure}
\epsscale{0.75}
\plotone{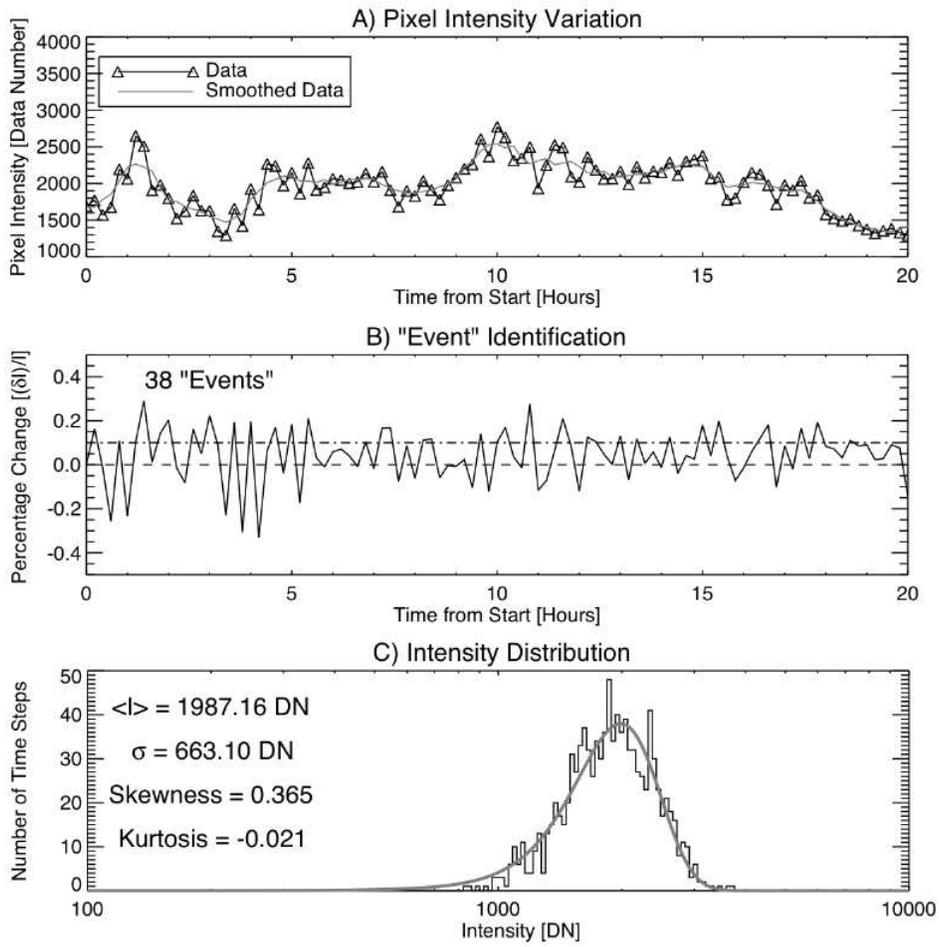}
\caption{Definition of the basic timeseries diagnostics used in this analysis (see text for details).\label{diag}}
\end{figure}

\begin{figure}
\epsscale{1.0}
\plotone{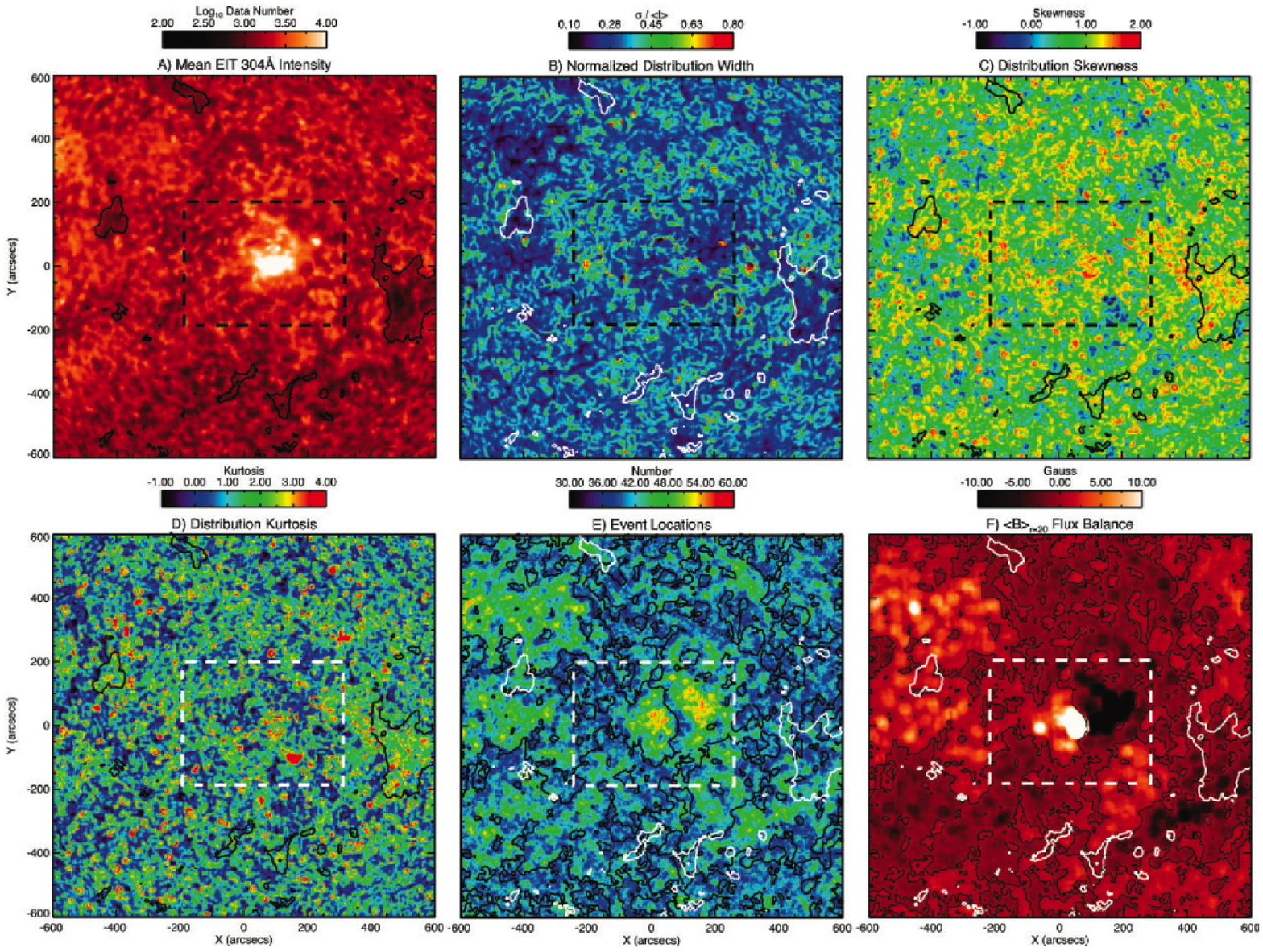}
\caption{Distribution distribution maps for the March 2006 sequence. From left to right and top and bottom rows we show the mean intensity, normalized intensity distribution width, skewness, kurtosis, event location and \brr{} maps. In each panel we show the EIT 195\AA{} 150DN intensity level while, in panels E and F, we show the magnetic neutral line (\brr=0G). The dashed rectangle isolates the active region shown in Fig.~\pref{arr}. The online edition of the Journal has a version of this figure that shows the watershed segmentation boundaries derived from the 304\AA{} intensity image of panel A and also provides a movie of the entire 304\AA{} timeseries analyzed. The 500\arcsec x 400\arcsec{} region around the active region (dashed rectangle) studied in this paper is shown for reference. \label{resultsmarch2006}}
\end{figure}

Applying the diagnostics discussed above to the EIT 304\AA{} timeseries sample we can develop an understanding of the relentless nature of magnetoconvection driven energy release in the transition region and lower solar corona through the transient brightenings that it produces. In Figure~\pref{resultsmarch2006} we present the diagnostic maps for the EIT sequence. Panels A through F show the mean EIT 304\AA{} intensity, the normalized width of the intensity distribution (width /  mean intensity), the intensity distribution skewness, distribution kurtosis, the number of ``events'' and the \brr{} respectively. In each panel we overplot the EIT 195\AA{} iso-intensity contour at 150DN (e.g., Fig.~\pref{march2006}A) while, in panel F, we show the magnetic neutral line as a thin black contour (e.g., Fig.~\pref{march2006}D). The online version of the Journal includes a version of this figure which includes the watershed segmentation supergranular boundaries \cite[e.g.,][]{Lin2003} that are derived from the mean 304\AA{} intensity over the period studied. The 500\arcsec x 400\arcsec{} rectangle around the active region (dashed rectangle) is shown for reference.

Panel B shows the normalized width of the intensity fluctuation in each sequence. We see that the net width of the intensity distributions are almost uniformly reduced in the coronal holes on-disk (inside the EIT contours) where the mean intensity is also much smaller. There is a prevalent pattern of ``rings'' in this panel that appear to be of a size commensurate with the supergranular network, visible in panel A. In addition, there are positions of very large normalized widths (``hot spots''; at positions such as [-250\arcsec,-325\arcsec], [-300\arcsec,50\arcsec]) which are co-spatial with another set of rings that are clearly visible in panels C (skewness) and D (kurtosis). The skewness and kurtosis rings underlie the positions of EUV Bright Points (BPs) in the 195\AA{} image (McIntosh 2007 in preparation). Filament channels present their own particular signature in the normalized width maps (see, e.g., [-600:-200\arcsec, -400:0\arcsec]) as nearly continuous lines (``snakes'') of low distribution width in the center of the filament outlined with piecewise continuous regions of large distribution width. Active and plage regions (e.g., [0:200\arcsec, -100:100\arcsec]) are most easily identified in panel E when the number of brightening events detected increases almost 75\% above the mean background value ($\sim$35). 

Overall, we see that there is a striking correspondence between the number of events counted (panel E) and the variation of the \brr{} (panel D). From the discussion in Sects.~4 and~5 of \cite{McIntosh2006b} we would expect this to be the case if magnetoconvection-driven reconnection were responsible for their appearance; flux emerging into a region of larger imbalance would increase the probability of immediate reconnection. The same connection to patterning induced by magnetoconvection-driven reconnection can be attributed to the quiet sun distribution width rings. They occur almost exclusively at the interior of the supergranular boundaries and bear an uncanny resemblance to the loci of highest reconnection probability in the cartoon representation of the reconnection process \cite[Fig~12 of][]{McIntosh2006b}. In this Letter we focus on the intensity diagnostic patterns observed in the vicinity of an active region and offer a new insight into the mass and energy loading of active regions from their base, leaving other features for a future publication (McIntosh, Judge \& Gurman 2007 - in preparation).

In what follows, we assume that the emission in the 304\AA{} bandpass, outside of the brightest active regions pixels, is dominated by \ion{He}{2}; we believe this assumption is justified both by the response of the EIT multilayer coatings \cite[][]{Boudine1995} and spectroscopic measurements of quiet and active Sun intensities \cite{Brosius1998}. Similarly, we assume that the constant driving of the emerging magnetic flux by the convective flow in the high plasma-$\beta$ will not switch off unless the flow itself is suppressed, i.e., in a sunspot spanning at least one or two supergranular cells in diameter. From additional context images supplied on the \soho{} website ({\url http://sohowww.nascom.nasa.gov/cgi-bin/get\_soho\_images?summary+20060319}) or on the Active Region Monitor \\  ({\url http://www.solarmonitor.org/index.php?date=20060319}), there is a complex of very small sunspots underlying the active region, but that these spots are not of sufficient size to impede the flow significantly outside the spots themselves.

\begin{figure}
\epsscale{0.75}
\plotone{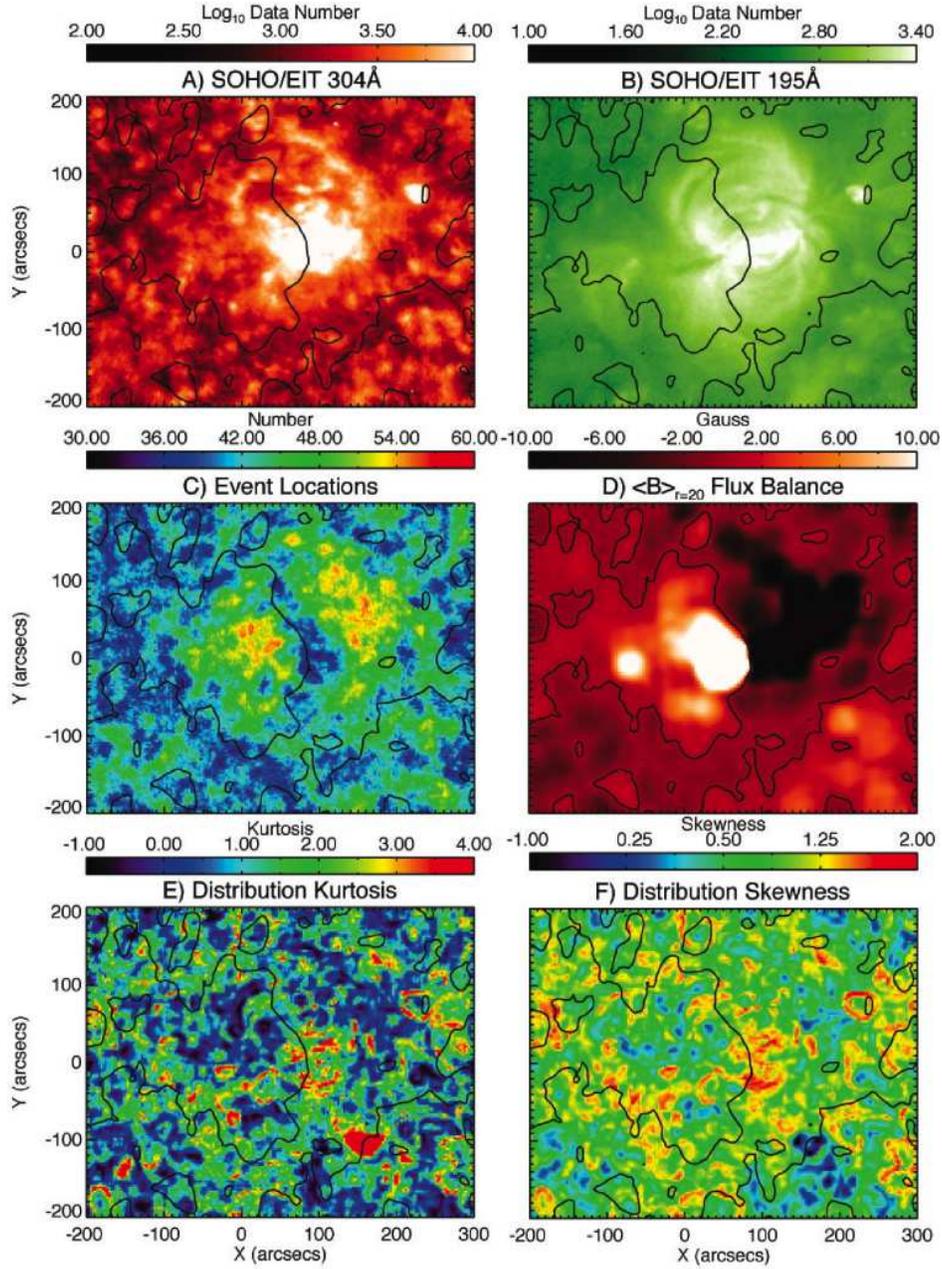}
\caption{An expanded view of the active region present in the March 2006 sequence (cf. Fig.~\pref{resultsmarch2006}). \label{arr}}
\end{figure}

Figure~\pref{arr} shows the maps of the intensity distribution diagnostics in and around a relatively simple looking active region complex [NOAA ARs 10860 and 10862]. The panels of Fig.~\pref{arr} show the mean EIT 304\AA{} intensity, 195\AA{} image, event location and the \brr{} maps for the active region. We see that spatial distribution of events (panel C) shows two lobes of intense activity where the number of events increases considerably ($\sim$60 events per pixel) over that of the quiet sun ($\sim$45 events per pixel) and the supergranular cell interiors in particular ($\sim$35 events per pixel). The event pattern is visibly correlated to the structure and amount of imbalance in the opposing polarity lobes of the active region itself with a measurable drop overlying the \brr{} neutral line (panel D) despite the bright emission over the same location in panels A and B. The location of the events tell us a great deal about how mass and energy are transported into the active region and plage complexes. This correspondence may also shed light on why neighboring coronal loops (anchored in the same active region) have apparently different mass loads and temperatures \cite[e.g.,][and most TRACE observations]{Klimchuk2006}. Panels E and F display enhanced distribution skewness and kurtosis in regions that bisect the major magnetic neutral line [100\arcsec, 0\arcsec]. We speculate that this, highly impulsive, but small (not a large number of events) magnitude intensity change in the bright emission above the sunspots has a significant contribution from the coronal emission contaminating the the EIT 304\AA{} passband (see the movies accompanying Fig.~\pref{resultsmarch2006}), but acknowledge that information is really limited and this must be verified by modeling and more detailed observations in the passband, such as discussed by \cite{JudgePietarila2004,Pietarila2004} and those from the EUNIS sounding rocket \cite[][]{Thomas2001}.

\section{Discussion \& Conclusion}

If we assume that we are indeed observing the impulsive mass loading and thermal heating caused by magnetoconvection-driven ejecta that are trapped below the coronal loop arcade \cite[an extreme case of the quiet Sun heating and mass loading presented][]{McIntosh2006b}, then the most activity takes place in the largest regions of \brr{}. We would expect this to be the case when the emerging flux is almost immediately forced to reconnect in those locations because the net reconnection rate must increase as \brr{} increases. Similarly, the magnitude of the reconnection event (the energy input into the system) will be greater at those locations, as there as there is a larger reservoir of magnetic imbalance to exploit, the drop in the event production rate around the neutral line to the quiet sun value appears to add weight to this argument. The increased variation of the field inclination in the evolving high-$\beta$ magnetic field will also result in an increased ``leakage'' of {\it p-modes} in and around the active region \cite[][]{Jefferies2006} that contain an energy flux may well dwarf that of the forced reconnection that initiated their propagation along the field lines. 

The net result of the footprint dominated energy release discussed here is that we will observe coronal loops within the active region (and plage) complex where there is contrast in the thermal and mass load based on the contrast in the magnetic flux available where they are anchored. Further, we believe that the hotter (low-$\beta$) corona may well be sustained by an entirely different physical energy release \cite[e.g., nanoflaring][]{Parker1988} where the energy (and mass) is subsequently (and efficiently) redistributed by thermal conduction along the lines of force. In this case we have a ``two stage'' heating process for the active Sun \cite[as in the scenarios advocated by][]{Klimchuk2006, Aschwanden2007}. There is a dominant hydrodynamic forcing of the plasma that initiates the mass and energy loading from the chromosphere and a secondary process that acts to redistribute the small percentage of the mass and energy along the magnetic fibers that compose the low-$\beta$ portion of the active region. The second stage, intrinsically tied to individual lines of force and their interaction, should help resolve the apparent spatial thickness of low\--$\beta$ coronal loops \cite[][]{Lopes2006,DeForest2006}. %The mass and energy loading in the chromosphere ultimately determines the structure observed in the corona and the division of the solar plasma into high\-- and low\--$\beta$ regimes is more natural (physically-based) than the prevalent (observationally\-- and historically-based) photosphere, chromosphere, transition region and corona.

Further, the results presented here may offer some perspective on the anomalously high UV/EUV helium emission observed on the Sun compared to theoretical models \cite[][]{Jordan1975}. We note that the combination of magentoconvection-forced reconnection ejecta and leaking {\it p-modes} in a class of chromospheric spicule provide evidence in support of \cite{JudgePietarila2004} that spicule forced neutral helium diffusion across the chromospheric magnetic topology is the significant contributor to quiet Sun helium emission. In addition, \cite{JudgePietarila2004} argue that the influence of the ``photoionization-recombination'' mechanism \cite[][]{Zirin1975} of overlying coronal radiation is negligible for the \ion{He}{2} 304\AA{} emission that is formed in the hybrid physical region at the boundary of the chromosphere and transition region. \cite{Mauas2005} come to the same conclusion for active region UV/EUV helium emission. The other proposed helium emission enhancement mechanism, ``velocity redistribution'' \cite[from non-thermal plasma motions in the line formation region, e.g.,][]{Andretta2000}, is the result of ``transient ionization'' \cite{Pietarila2004}. We ask if it possible that this rapid transient ionization is a result of impulsively heating of the plasma as would happen locally in the forced reconnection ejecta we study in this paper? This question will be addressed at greater length in McIntosh, Judge \& Gurman (2007).

%\section{Conclusion}

To conclude, We have offered an interesting insight into the dynamic loading of (active region) coronal loop arcades from the high\--$\beta$ region in which they are tethered. In fact, we observe that the relentless loading is driven by the same mechanism responsible for the quiet Sun and coronal holes, while these have a clear topological difference on the global scale (``open'' versus ``closed'' flux) an active region highlights the degree of energy input into the system from its footprint; we have seen that it scales as the degree of imbalance in the local magnetic field grows. %We propose that this result may have serious consequences for coronal mass ejections (CME) and the post-CME evolution of the coronal plasma in its wake \cite{McIntosh+others2007a}. 
Unfortunately, we cannot use the data presented in this Letter to discuss the exact micro-physics of how the dynamically forced loop footpoints in the active region transport their mass and energy into the structures observed in the hotter (low-$\beta$) coronal plasma to answer, unambiguously, the question posed in the title of this Letter, but the detailed observations to be made by {\it Hinode} and EUNIS \cite[e.g.,][]{Brosius2007} will go a long way to resolving this picture.

\acknowledgements 
The author would like to thank Tom Bogdan, Alisdair Davey, Joe Gurman, Phil Judge and Meredith Wills-Davey for very helpful discussions. 
The material presented is based upon work carried out at the Southwest Research Institute supported by grants from NASA (\soho{} - NNG05GQ70G; SEC-GI - NNG05GM75G; SR\&T - NNG06GC89G) and the NSF (ATM-0541567) issued to the author.

\clearpage

%In panels A through D we show the EIT 195\AA{} coronal image, the first EIT 304\AA{} image in the sequence, the MDI $B_{||}$ image and supergranular-scale smoothed $B_{||}$ image \cite[\brr;][]{McIntosh2006a}. In each panel we show the EIT 195\AA{} 150 Data Number (DN) contour (white) and the 1200\arcsec x 1200\arcsec{} field of view studied (black square in panels A and B, white in panels C and D). In panel D the thin black contour shows where \brr=0~G, the magnetic neutral line. 

\clearpage

%Definition of the basic timeseries diagnostics used here. Panel A shows the intensity lightcurve (joined triangles) in the sample pixel with the box-car smoothed intensity trend (thin gray line) that is required for panel B to identify 10\% changes (horizontal dot-dashed line), or brightening ``events'' in the trend-removed intensity. Panel C shows the intensity distribution for the pixel, provides the four moments of the sample distribution and shows the result of a Gaussian fit to the distribution (thick gray line). 

\clearpage

\clearpage

%We show the mean EIT 304\AA{} intensity (panel A) as well as the 195\AA{} EIT context (B), the event location (C), \brr{} (D), distribution kurtosis (E) and skewness (F) maps from top to bottom and left to right respectively. Each panel shows the contour of \brr=0G. 

\end{document}